\documentclass[pra,twocolumn,showkeys,preprintnumbers,amsmath,amssymb]{revtex4}
\makeatletter
\usepackage[dvips]{graphicx}	

\usepackage{amsmath}
\usepackage{amssymb}
\usepackage{amsbsy}
\usepackage{color}
\setcitestyle{square}
\usepackage{epstopdf}
\DeclareGraphicsExtensions{.pdf,.eps,.png,.jpg,.mps}

\begin{document}

\title{
Metal-semiconductor behavior along the line of stacking order change \\ in gated multilayer graphene}
\author{ W. Jask\'olski} 
\email{wj@fizyka.umk.pl}
\affiliation{Institute of Physics, Faculty of Physics, Astronomy and Informatics, Nicolaus Copernicus University in Torun, Grudziadzka 5, 87-100 Toru\'n, Poland}

\begin{abstract}
We investigate gated multilayer graphene with stacking order change along the armchair direction. We consider some layers cracked to release shear strain at the stacking domain wall. The energy cones of graphene overlap along the corresponding direction in the $k$-space, so the topological gapless states from different valleys also overlap. However, these states strongly interact and split due to atomic-scale defects caused by the broken layers, yielding an effective energy gap. We find that for some gate voltages, the gap states cross and the metallic behavior along the stacking domain wall can be restored. In particular cases, a flat band appears at the Fermi energy. We show that for small variations of the gate voltage the charge occupying this band oscillates between the outer layers.
\end{abstract}


%\date{\today}

\maketitle

\section{Introduction}

Stacking order domain walls in multilayer graphene are most usually considered along the zigzag direction. They can be created when some layers are stretched, corrugated, or delaminated along the armchair direction \cite{Ju_Nature_2015,Pelc_2015,Lane_2018,Anderson_PRB_2022}. Stacking order change causes the appearance of topological states. When the multilayer is gated and the gap opens, the one-dimensional topological gapless states appear at valleys $K$ and $K^{\prime}$ separated by $\frac{4}{3}\pi/a$, \cite{Vaezi_2013,Zhang_Pnas_2013,Alden_2013,San_Jose_2014,Jaskolski_2020,Ju_Nature_2015}. The gapless states are then valley-protected as long as there is no valley mixing caused, for example, by atomic-scale defects. It was however shown that even under strong defects like bonds breaking in some layers at the domain wall, some gap states that connect the valence band (VB) and conduction band (CB) continua persist, and the systems preserve metallic character along the stacking domain wall \cite{Jaskolski_2019,Jaskolski_2021,Jaskolski_2023}. 

The situation is different if we allow stacking domain wall along the armchair direction. Firstly, because along the armchair direction, the valleys appear at the same $K=0$ \cite{Yin_NC_2016}. Therefore, the gapless states, which have opposite slopes $E(k)$ at each valley, overlap and may no longer be valley-protected. Secondly, because such stacking change causes strong shear strain at the interface between different stackings, at least for narrow domain walls.

%% FIGURE %%%
\begin{figure}[ht]
\centering
\includegraphics[width=\columnwidth]{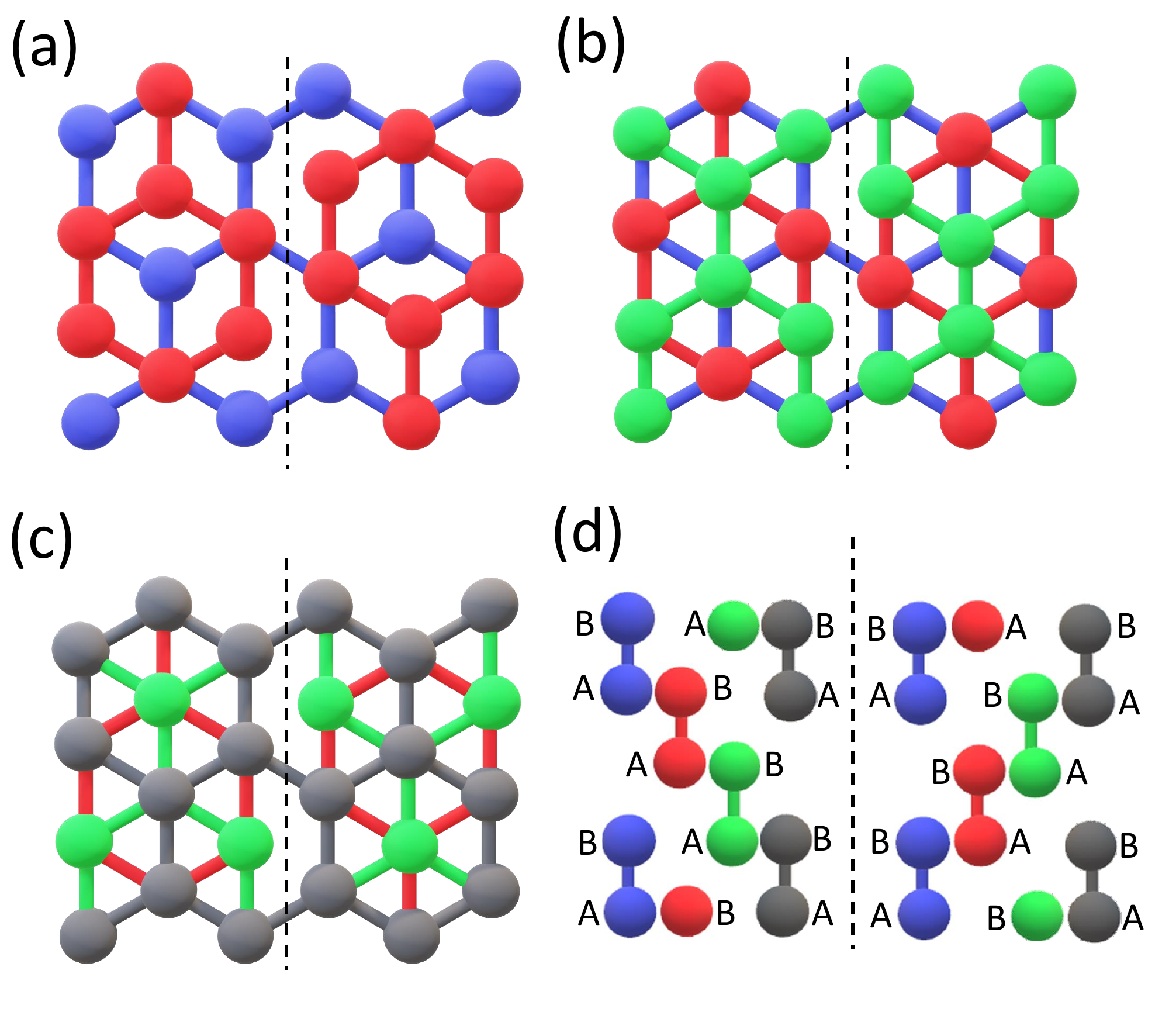}
\caption{\label{fig:first}
Schematic representations of the investigated systems. 
(a) BLG with stacking order change AB/BA. (b) TLG with stacking orders ABC/BAC. (c) FLG with stacking order change ABCA/BACB. (d) Lateral view of sequences of carbon atoms in the consecutive layers along the armchair direction. Vertical broken lines mark the interface between different stackings. The layers from bottom to top are marked in colors: blue, red, green, and dark grey. The bottom (blue) layer in (c) is covered and not seen.
}
 \end{figure}

In this work, we investigate the behavior of gapless states in such systems. We consider multilayers in which the interface bonds are broken in some layers to release any shear strain in the system. Fig. \ref{fig:first} shows how the unstrained bilayer (BLG), trilayer (TLG), and four-layer graphene (FLG) with stacking domain walls along the armchair direction can be formed. One can easily imagine that such cracks of layers may be present in realistic multilayers \cite{Cracks_Beitner_NL_2023, Jang_Nanoscale_2017}. The bonds broken in some layers act as atomic-scale defects. Therefore, the gapless states strongly interact and split yielding the appearance of the effective energy gap and thus semiconducting properties of such multilayers. However, we show that for some gate voltages, some gap states cross and the energy gap disappears. The metallic behavior along the stacking domain wall can be restored. Transitions from semiconductor to metal can be achieved by changing the gate voltage. Moreover, for some gate voltages, the separated gap states in FLG connect at the Fermi level ($E_F$) leading to a kind of flat band at $E_F$, with charge occupying this band localized mainly in the outer layers.

\section{System and method}

We investigate Bernal stacked bilayer and rhombohedral four-layer graphene with stacking order domain walls. Fig. \ref{fig:first} demonstrates how the consecutive layers are deposited. Panels (a) and (c) show the investigated BLG and FLG systems, and panel (b) shows an intermediate system of the TLG. Only small fragments of the multilayers are shown, while the investigated systems are infinite in both in-plane directions. The bottom layer is always a uniform graphene, as well as the top layer in the case of FLG. The nodes of the FLG top layer have the same in-plane positions as the nodes of the bottom layer. The remaining layers are broken, they consist of two nonconnected half-infinite graphene layers shifted relative to each other by $a_{C-C}$ along the armchair direction. This allows for the formation of a strainless interface with different stacking orders of rhombohedral multilayers on both sides. Note that due to the presence of the stacking domain wall, the systems studied are periodic only in the direction defined by the domain wall, i.e., in the armchair direction. Fig. \ref{fig:first} (d) shows the stacking order of layers on both sides of the domain wall. For BLG they are simply Bernal stackings AB and BA, respectively, while for FLG we have ABCA and CBAC stackings. 

To investigate the electronic structure of the defined multilayers we calculate the local density of states (LDOS) at the interface. The calculations are performed using the Green function matching technique for a system composed of a conductor connected to two semi-infinite leads \cite{Nardelli_1999}. In our case, the conductor consists of the interface region, as defined in Fig. \ref{fig:first} (a)-(c), connected on the left- and right-hand sides to two semi-infinite multilayers. The local density of states is calculated as ${\rm{LDOS}}(E)=-\frac{1}{\pi}{\rm{Im[Tr G_C}}(E)]$, where ${\rm{G_C}}$ is the interface Green function determined by the Hamiltonians of the conductor and the leads. The Hamiltonians are calculated using a tight-binding approximation. The values of intra-layer and inter-layer hopping parameters are $t_i=2.7$ eV and $t_e = 0.27$ eV, respectively, following the \cite{Castro_2007,Ohta_2006,Jaskolski_2016,Jaskolski_2D}. Since the system is periodic in the armchair direction, the local density of states is $k$-dependent, i.e., LDOS($E$,$k$), where $k$ is the wave vector corresponding to this periodicity.

%% FIGURE %%%
\begin{figure}[ht]
\centering
\includegraphics[width=\columnwidth]{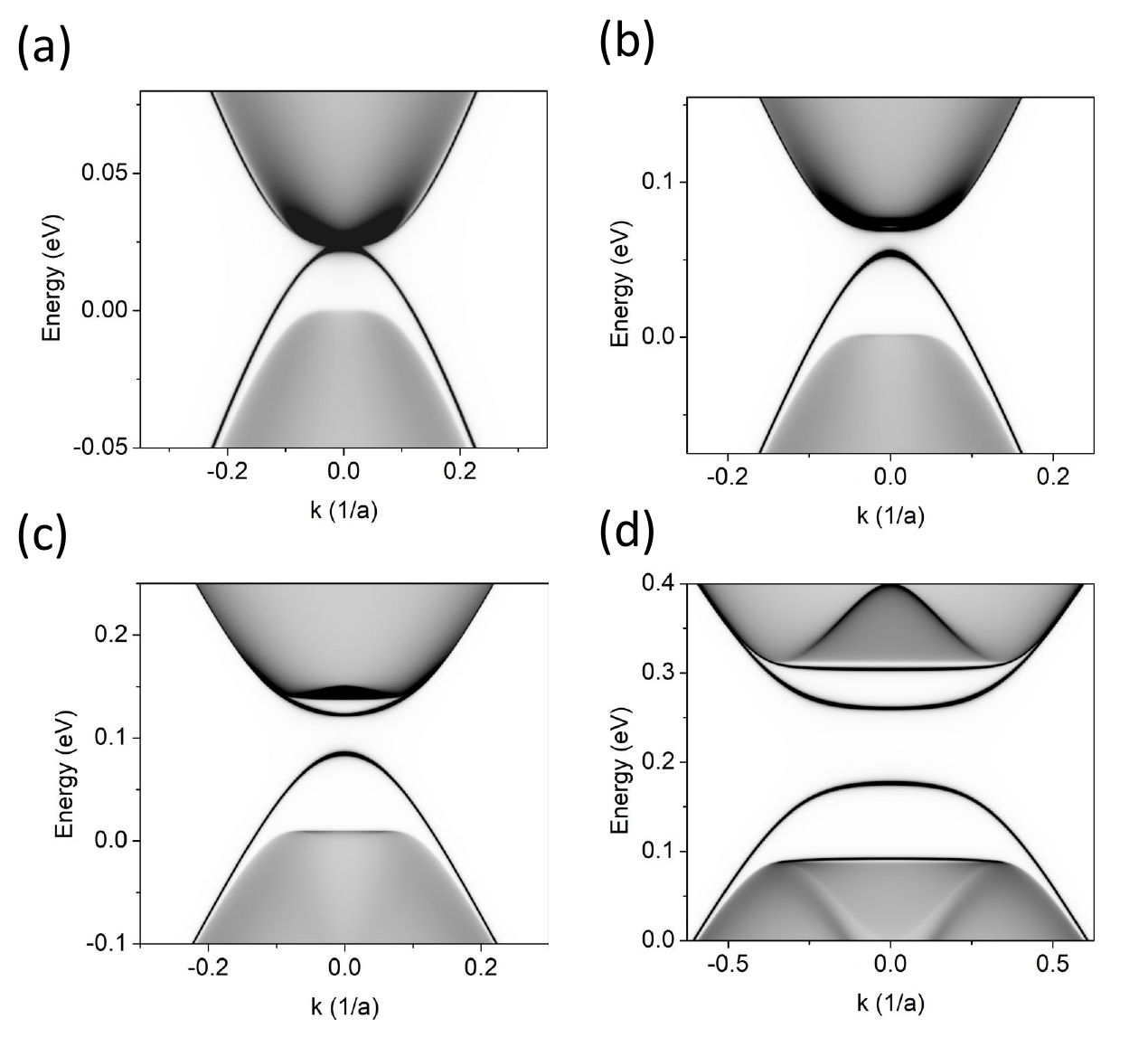}
\caption{\label{fig:second}
LDOS for BLG with stacking order domain wall along the armchair direction, as visualized in  Fig. \ref{fig:first} (a), for different gate voltages $V$ applied between the layers. (a) $V=0.025$ eV, (b) $V=0.075$ eV, (c) $V=0.15$ eV, (d) $V=0.4$ eV. The Fermi level lies in the middle of the gap defined by the energy continua. 
}
 \end{figure}

%% FIGURE %%%
\begin{figure}[ht]
\centering
\includegraphics[width=\columnwidth]{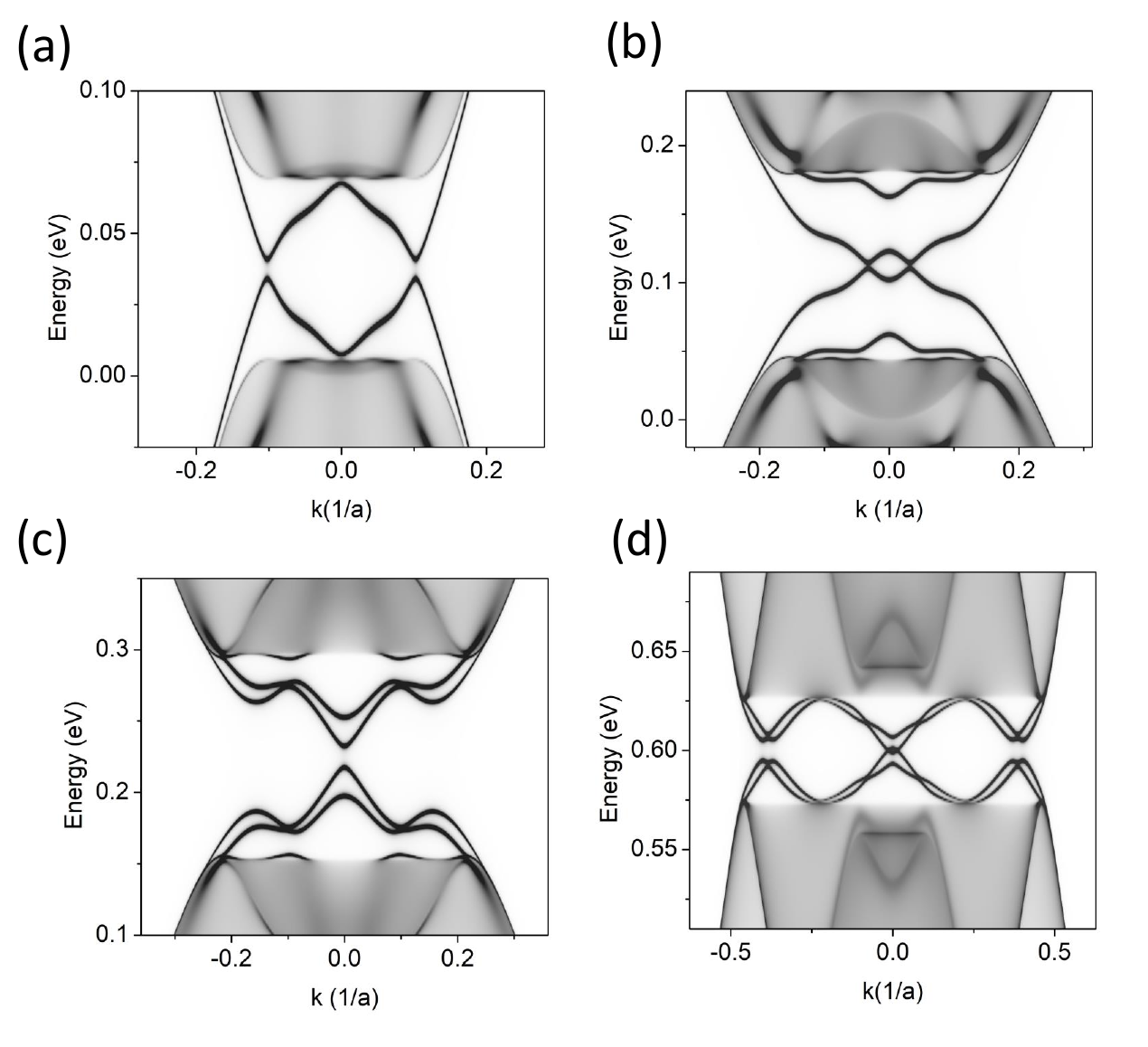}
\caption{\label{fig:third}
LDOS for FLG with stacking order domain wall along the armchair direction, as visualized in  Fig. \ref{fig:first} (c), for different gate voltages $V$. (a) $V=0.025$ eV, (b) $V=0.075$ eV, (c) $V=0.15$ eV, (d) $V=0.4$ eV. The Fermi level lies in the middle of the energy gap. 
}
 \end{figure}

\section{Results and discussion}

We perform calculations for defined graphene multilayers in a perpendicular external electric field. Several values of the field are considered, namely, {\bf{E}}=7, 22, 44, and 120 mV/\AA. Assuming the interlayer distance $d=3.35$ {\AA} the values of {\bf{E}} translate into gate voltages between the neighbor layers $V=0.025$, 0.075, 0.15, and 0.4 eV. The voltages we consider are available in the experiment and are of the same order as typically used for multilayer graphene systems \cite{Jaskolski_2D,Li_NN_2016,Sena_Peeters_PRB_2014,LC_Carbon_2024}.

In Fig. \ref{fig:second} we show LDOS calculated in the BLG system for four values of the gate voltage $V$. Of the four gapless states that may appear at all at $K=0$ in the BLG with a stacking domain wall \cite{Yin_NC_2016}, there is little left in the energy gap. For $V=0.4$ eV, four separated flat bands are visible as a remnant of interacting topological gapless states (two of them are situated very close to the band continua). This is caused by the atomic-scale crack in the upper layer. For smaller $V$, fewer bands are visible in the energy gap. Finally, for $V=0.025$ eV, the band that "grows up" from the VB continuum connects with the conduction band and the system takes on a metallic character along the domain wall.  

Figure \ref{fig:third} shows LDOS for FLG system. Now, the situation is more complex. The stacking order change may, in principle, lead to even eight gapless states at $K=0$, but here the number of bands in the energy gap depends strongly on $V$. For very small $V=0.025$ eV the overall picture resembles two pairs of intersecting topological states with opposite slopes. They slightly repeel at the crossing points, so a small gap opens at the $E_F$. For higher values, e.g., $V=0.075$ eV, this small gap closes, the band crossing points approach the cone center and two new bands detach from the VB and CB continua. Increasing farther the gate, the gap opens (for $V\approx 0.1$ eV), but closes again for $V$ higher than 0.3 eV and the system restores again its metallic character.

\begin{figure}[ht]
\centering
\includegraphics[width=\columnwidth]{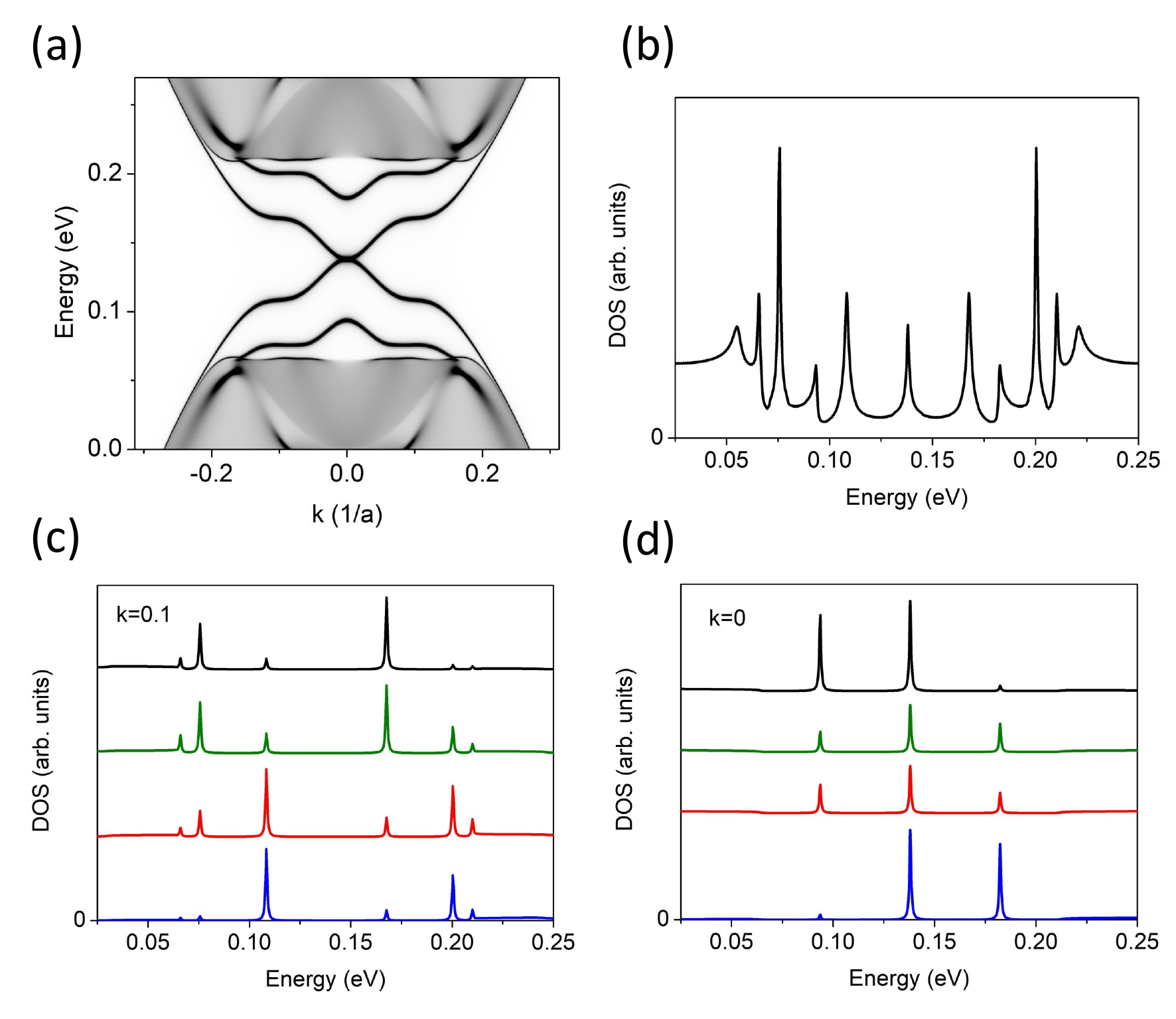}
\caption{\label{fig:fourth}
(a) LDOS for FLG and $V=0.092$ eV. (b) PDOS, i.e., LDOS integrated over $k$ in the IBZ.
(c) and (d) LDOS distribution in the layers for $k=0.1$ and $k=0$, respectively. The density in consecutive layers, from bottom to top, is shown in colors: blue, red, green, and black.
}
 \end{figure}

Especially interesting is the case when $V\approx 0.092$ eV. The corresponding LDOS is presented in Fig. \ref{fig:fourth} (a). The four bands in the energy gap show flat areas and a short flat (in the range of $k$) band is formed also at the Fermi level where two bands touch. The increased density of this flat band  at $E=0.14$ eV is well seen in Fig. \ref{fig:fourth}, where the projected density of states, i.e., PDOS($E$)=$\int_{k\in IBZ}{\rm{LDOS}}(E,k)dk$ is shown. The distribution of band densities in the layers is visualized in Fig. \ref{fig:fourth} (c) and (d). Of the two bands that touch at the Fermi level, the band below the $E_F$ is localized in two lower layers, while the band above the $E_F$ localizes in the two upper layers. Slightly away from the cone center, i.e., for $k=0.1$, the distribution over layers is uniform. However, for $k=0$, i.e., just at the flat band, a fraction of the charge moves to the outer layers. The similar effect of leakage of charge to the outer layers was recently observed also in \cite{LC_Carbon_2024} for a system of double twisted trilayers.

\begin{figure}[ht]
\centering
\includegraphics[width=\columnwidth]{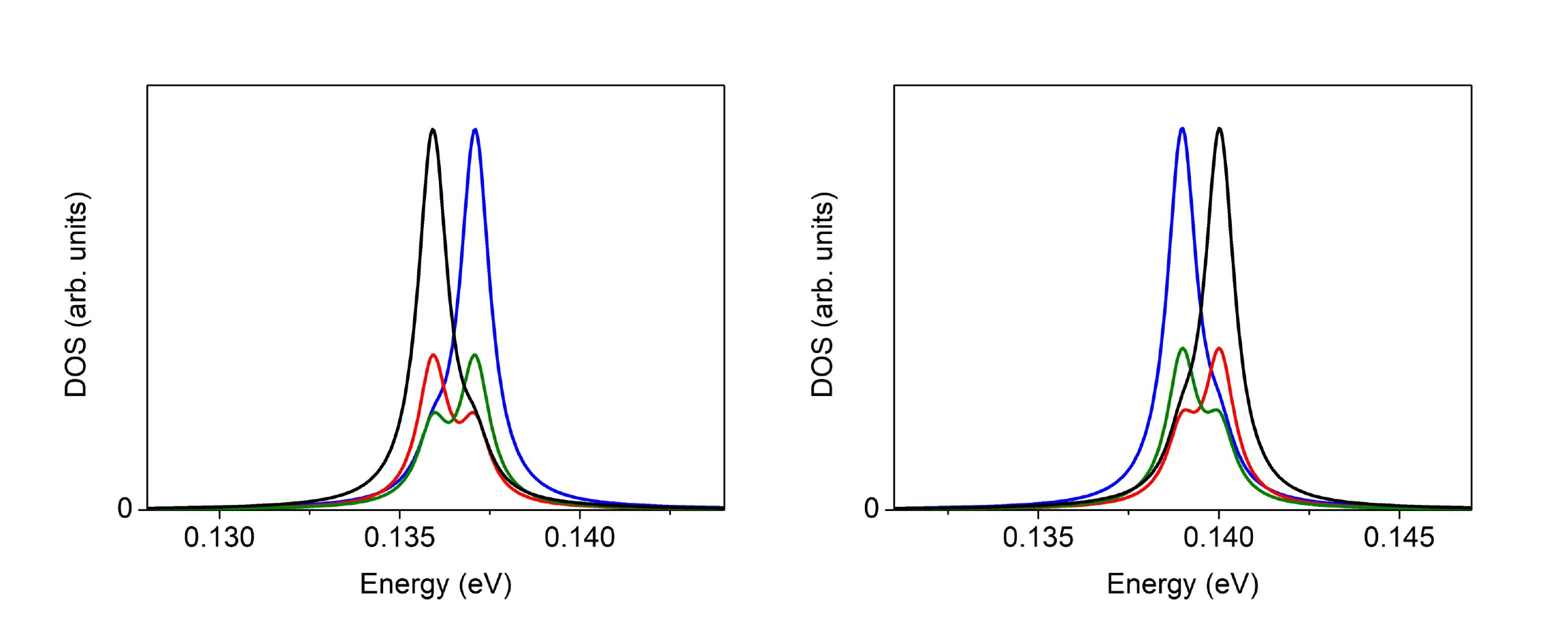}
\caption{\label{fig:fifth}
The LDOS($E$,$k=0$) distribution in the layers for FLG and two different gate voltages $V$. Left - $V=0.091$ eV, right - $V=0.093$ eV. The density in consecutive layers is shown in colors: blue, red, green, and black.
}
 \end{figure}

In low temperatures the flat band at $E_F$ is only half-filled. Note, that in graphene systems such a band can give rise to magnetic effects \cite{Kusakabe_book,Bouzerar_PRB_2023,Sawada_APL_2014,Pons_PRB_2020} and superconductivity \cite{Cao_N_2018,Han_NN_2023,Yin_PRB_2020,Chittari_PRL_2019}.  The width of the flat band at $E_F$ is about 5 meV. In Fig. \ref{fig:fifth} we show the LDOS($E$,$k=0$) of this band resolved into the layers, for two values of the gate voltage, $V=0.091$ eV and $V=0.093$ eV, for which the falt band is still present at $E_F$. This result confirms that at $k=0$ the density is localized mainly in the outer layers (blue and black). More importantly, we find that a slight increase of $V$ by only a few mV, moves the charge occupying the flat band at $E_F$, from the top to the bottom layer.

%% FIGURE %%%

%% FIGURE %%%

The calculations have been also performed for trilayer (Fig. \ref{fig:first} (b)), but since the results do not differ substantially from those for BLG, they are not presented. We have also checked that wider separation of the half-graphene layers does not essentially change the results. The band crossing points and the separation of bands can occur for different voltages, but the overall picture of the gap states and their behavior with the changing $V$ is similar.

\section{Conclusions}

We have studied the electronic structure of gated bilayer and four-layer graphene with stacking order domain walls created along the armchair direction. Bernal and rhombohedral stackings have been considered for BLG and FLG, respectively. Since the change of stacking across the armchair direction causes shear strain at the domain wall, we allow some layers to crack, i.e., to break C-C bonds along the domain wall to release the strain. 

We have investigated the local density of states along the domain wall, i.e., at the interface of the adjacent stacking orders. The graphene energy cones $E(k)$ overlap in the corresponding direction in the $k$-space. Thus, all the possible topological gapless states appear at the same valley, $K=0$. Since the broken bonds act as atomic-scale defects and lead to the removal of valley protection of gapless topological states, these 
states interact and split yielding an effective energy gap. We have, however, found that for some gate voltages, the metallic character of the multilayer along the domain wall is restored. In the case of BLG, it happens for very small gates. In the case of the four-layer graphene, the gap states cross or anticross, i.e., they close or open the energy gap, depending on the gate. 

In the case of four-layer graphene and for a specific value of the gate voltage, some bands in the gap connect forming a kind of a short flat band at the Fermi energy. We have demonstrated that the density of this band is moved to the outer layers. Finally, we have also shown that for very small changes in the gate voltage, the charge that fills the flat band at the $E_F$, can change its localization between the bottom and the top layers. 
This gives us a tool for changing the layer localization of one-dimensional currents that at $E_F$ can flow along the stacking domain wall.

\end{document}